\documentclass[aps,nofootinbib,twocolumn,longbibliography,prd,showpacs]{revtex4-1}
\usepackage{amstext,amsmath,amssymb,amsfonts,bbm}
\usepackage[latin1]{inputenc}
\usepackage{graphicx}
\usepackage{tocvsec2}
\usepackage{color}

\usepackage{multirow} \usepackage{rotating}
\usepackage[colorlinks,citecolor=blue,linkcolor=blue,urlcolor=blue]{hyperref}

\def\beq{\begin{equation}}
\def\be{\begin{equation}}
\def\ee{\end{equation}}
\def\bes{\begin{eqnarray}}
\def\ees{\end{eqnarray}}

\begin{document}
\maxtocdepth{subsection}
%%%%%%%%%%%%%%%%%%%%%%%%%%%%%%%%%%%%%%%%%%%%%%%%%%%

\title{\large \bf How to detect an anti-spacetime}
\author{{\bf Marios Christodoulou, Aldo Riello, Carlo Rovelli}}
\affiliation{Centre de Physique Th\'eorique, Case 907,  Luminy, F-13288 Marseille, EU}

\date{\small\today}

%%%%%%%%%%%%%%%%%%%%%%%%%%%%%%%%%%%%%
\begin{abstract}\noindent

\noindent Is it possible, in principle, to measure the sign of the Lapse?  We show that fermion dynamics  distinguishes spacetimes having the same metric but different tetrads, for instance a Lapse with opposite sign. This sign might be a physical quantity not captured by the metric. We discuss its possible role in quantum gravity.  \\[-1mm]

\noindent{\em Article awarded with an ``Honorable Mention" from the 2012 Gravity Foundation Award.}

\end{abstract} 
%%%%%%%%%%%%%%%%%%%%%%%%%%%%%%%%%%%%%%

%\pacs{04.60.Pp}
\maketitle

%%%%%%%%%%%%%%%%%%%%%%%%%%%%%%%%%%%%%%

\section{Negative Lapse}

Einstein discovered that the gravitational field can be described by a (pseudo-)Riemannian metric $g_{\mu\nu}(x)$.  But the Dirac equation, which governs the dynamics of fermions, cannot be coupled to this tensor. The solution is to replace $g_{\mu\nu}(x)$ by the tetrad field $e_{\mu}^I(x)$.\footnote{The tetrad is related to the metric by $g_{\mu\nu}(x)=\eta_{I\!J}e_{\mu}^I(x)e_{\nu}^J(x)$.  $\eta_{I\!J}$ is the Minkowski metric. Latin indices are 4d, raised and lowered with $\eta_{I\!J}$. Greek indices are tangent spacetime indices. We use also the differential-form notation $e^I=e^I_\mu dx^\mu$.} Using this field, the Dirac equation can be written in a general covariant form that describes the motion of fermions in a gravitational field. Since fermions exist, it makes sense to assume that gravity is  better described by the tetrad  than by the metric. Can this have direct observable consequences?  It is usually assumed that the answer is negative and that two fields $e_{\mu}^I(x)$ and  $\tilde e_{\mu}^I(x)$ defining the same metric are empirically indistinguishable. Here we suggest that this might not be the case. 

\begin{figure}[h]
\begin{center}
\centerline{\includegraphics[scale=0.5]{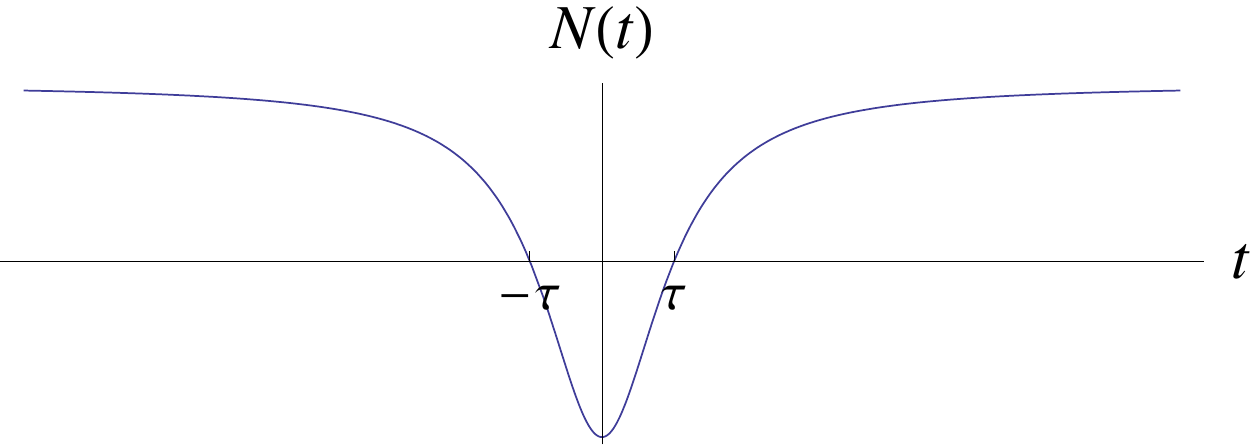}}
\caption{A Lapse function, becoming negative during a finite interval.}
\label{unoF}
\end{center}
\end{figure}

Consider a gravitational field given by:
\beq\label{uno}
 e^{i}= d x^{i},  \ \ \ \    e^{0}=N(t)dt.   
\ee
where $i=1,2,3$ and the Lapse function $N(t)$ is a smooth function which remains near unity everywhere but for a small region, where it takes negative value. As an example, consider for concreteness (see Fig.~\ref{unoF})
\be N(t)=\frac{t^2-\tau^2}{t^2+\tau^2}. 
\label{N}
\ee

\begin{figure}[h]
\begin{center}
\centerline{\includegraphics[scale=0.5]{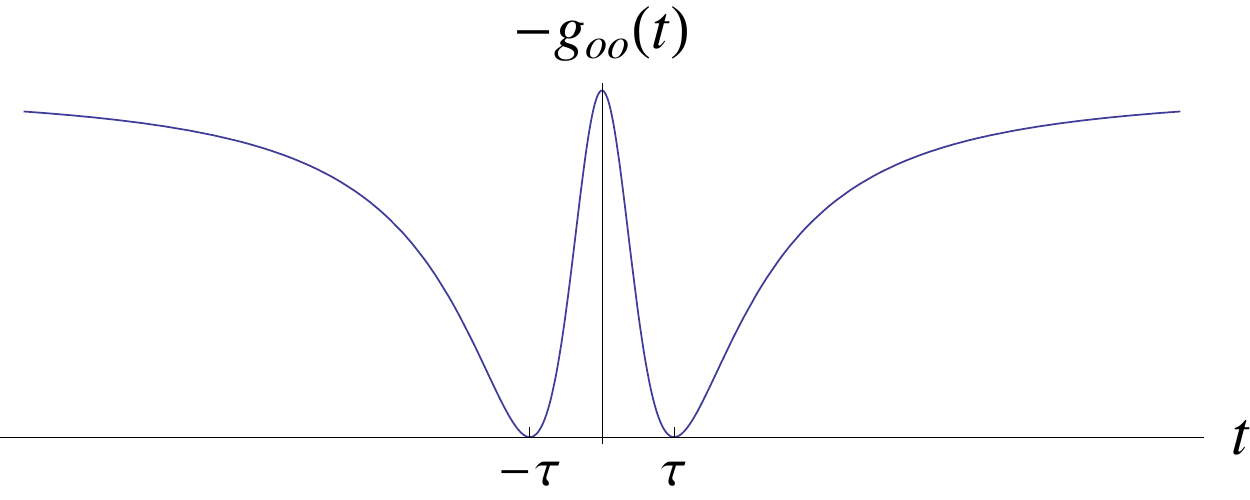}}
\caption{$-g_{00}(t)$ corresponding to the Lapse of Fig.~\ref{uno}.}
\label{g}
\end{center}
\end{figure}

\noindent The corresponding metric is $ds^2=-N^2(t)dt^2+d\vec x^2$. Of course $-g_{00}=N^2$ is positive; see Fig.~\ref{g}. 
\vspace{-2em}

This metric is easily seen to be the Minkowski metric in coordinates related to Minkowski coordinates $(T_M, \vec x)$ by $|N(t)|dt=dT_M$.\footnote{This coordinate transformation is $C^\infty$, but its inverse is not differentiable in two points. This is not worse than many far more singular coordinate choices routinely used  in general relativity.}  Therefore the gravitational field defined by \eqref{uno} seems physically equivalent to Minkowski space.  Is it really?  

\section{Measuring the sign of the Lapse}\label{secii}

Consider a massive fermion in a gravitational field  $e^I$(x). Its dynamics is governed by the Dirac equation
\be\label{dirac}
\gamma^I e_I^\mu \left[\partial_\mu+\frac{i}{2}\omega_\mu^{KL}J_{KL}\right] \psi+m\psi=0,
\ee
where $\gamma^I$ are Dirac matrices, $\omega_\mu^{IJ} $ is the torsionless, metric-compatible (Levi-Civita) spin connection determined by the tetrad, and $J^{IJ}\!=\!-\frac{i}{4}[\gamma^I,\gamma^J]$ are the generators of the Lorentz group in the fermion representation. 

Let the gravitational field be given by  \eqref{uno}. Its  spin connection vanishes. Assume for simplicity that the Dirac field is constant in space and choose a basis where $\gamma^0$ is diagonal.\footnote{This is the Dirac representation: $\gamma^0\!=\! {\scriptscriptstyle\left(\!\!\begin{array}{cc}-i\,\mathbb{I}& 0\\ 0&i\,\mathbb{I}\end{array}\!\!\right)}$ and $\gamma^i\!= {\scriptscriptstyle\!\left(\!\!\begin{array}{cc} 0 &-i\sigma^i\\ i\sigma^i&0\end{array}\!\!\right)}$,  where $\sigma^i$ are the  Pauli matrices.} Then a solution of \eqref{dirac} is 
\be
\psi(t)=e^{im\,f(t)}\psi_0 ,
\ee
where the time-dependent phase is given by 
\be
f(t)=\int N(t)\, dt \ \ =:\ \  T_{\!f}(t)
\label{phi}
\ee
and $(\psi_0)^A={}^T\!(1,0,0,0)$. We have introduced the notation $T_{\!f}$ for later convenience. Let's now have two such Dirac fermions $\psi_1$ and $\psi_2$, with different masses $m_1$ and $m_2=m_1+\Delta m$, and a device measuring the quantity $(\overline{\psi_1}\psi_2)$. Then, on the solution above we have
\be
(\overline{\psi_1}\psi_2)(t)=\exp[ i\phantom{,} \Delta m\phantom{,}  f(t)].
\ee
That is, the quantity $f(t)$ is observable in principle.  But $ f(t)$ depends on the \emph{sign} of the Lapse. It takes \emph{different} values on gravitational fields defining the \emph{same} Riemaniann geometry.  In other words, a fermion does not evolve in the gravitational field \eqref{uno} in the same manner in which it evolves, say, on the tetrad 
\be
\tilde e^0=|N(t)|dx^0,\ \tilde e^i=dx^i, 
\label{mt}
\ee
in spite of the fact that  the two tetrads define the same metric. In the presence of fermions, the sign of the Lapse is in principle observable.

\section{Fermion time} 

\begin{figure}[h]
\begin{center}
\centerline{
{\includegraphics[scale=0.3]{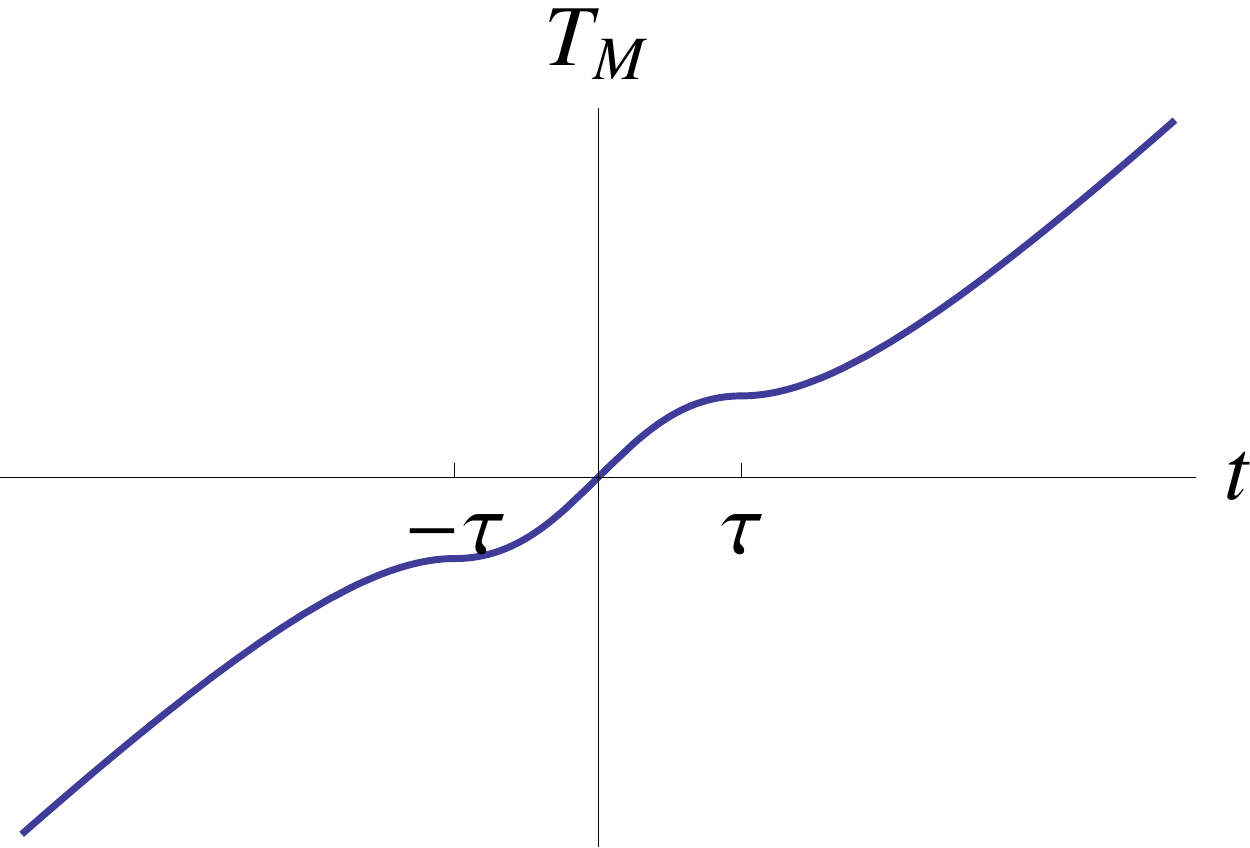} \hspace{.5em} \includegraphics[scale=0.3]{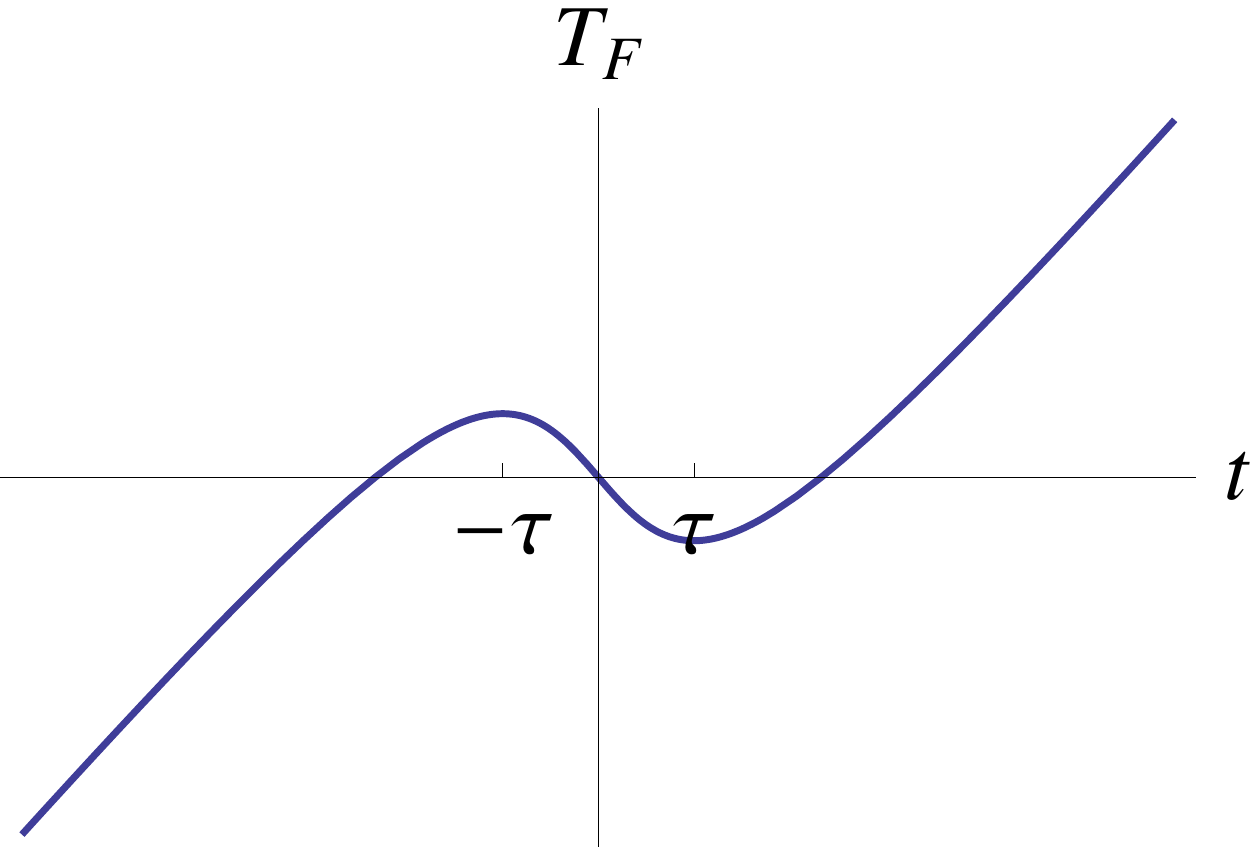}}}
\caption{Minkowski time (left) and fermion time (right), as a function of the coordinate time $t$, for the Lapse \eqref{N}.}
\label{tre}
\end{center}
\end{figure}

Let us return for simplicity to the gravitational field \eqref{uno}. A standard clock such as an harmonic oscillator, a pendulum, an atomic clock or an orbiting planet, measures  Minkowski time, which is related to the coordinate time $t$ by 
\be
 T_M(t)=\int \sqrt{-g_{\mu\nu}\frac{dx^\mu}{ds} \frac{dx^\nu}{ds}}ds=\int |N(t)|dt, 
 \label{tM}
\ee
plotted in the left Panel of Fig.~\ref{tre}. 

\begin{figure}[h]
\begin{center}
{\includegraphics[scale=0.4]{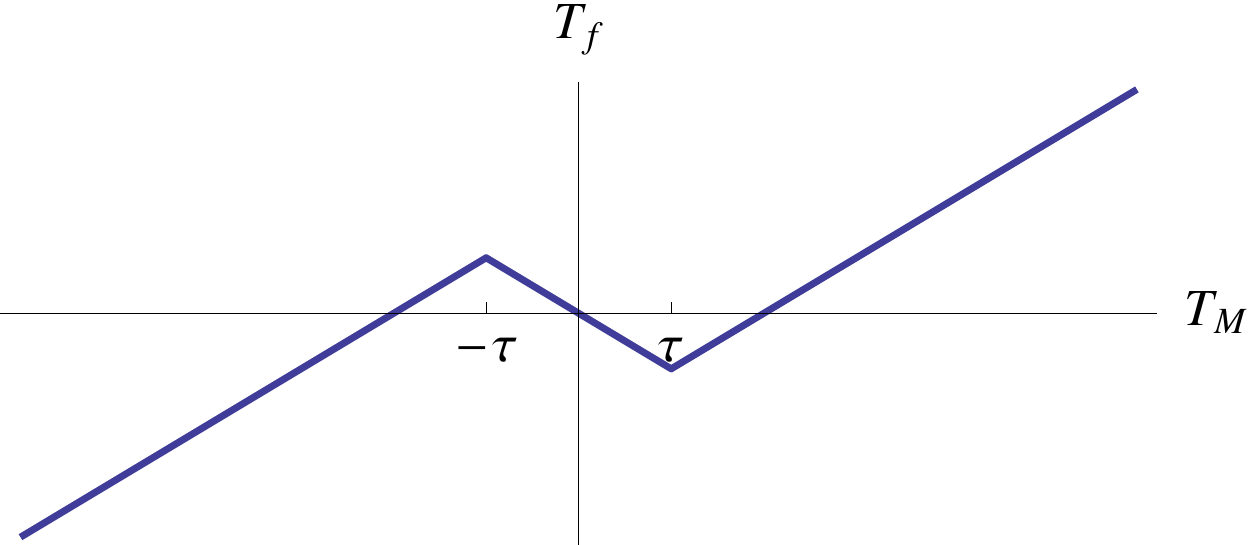}}
\caption{Fermion time as a function of Minkoswki time.}
\label{tf}
\end{center}
\end{figure}

As long as $N$ is positive, the phase  $T_{\!f}$ of the fermion defined in \eqref{phi} is  a good clock as well. Call  ``fermion time"  this quantity (hence the notation).  $T_{\!f}$ is plotted in the right Panel of Fig.~\ref{tre} as a function of the coordinate  $t$.  When  $N$ is negative, the fermion clock runs \emph{backward} with respect to the Minkowski time, as shown in Fig.~  \ref{tf}. Fermion time is observable, in spite of the fact that the metric field does not know about it.

A few comments are in order:
\begin{itemize} 
\item[i.]  Fig.~\ref{tf}  shows the non-differentiable relation between the times $T_{\!F}$ and $T_{M}$. Notice that both times are $C^\infty$ functions of $t$. The gravitational field \eqref{uno} is smooth, and can be defined on a $C^\infty$ manifold.  Minkowski time is a $C^\infty$ function on this manifold, but, since the inverse of $T_M(t)$ is non-differentiable in $t\!=\!\pm\tau$, the smooth structure defined by the Minkowski metric is not the same as the original one.  If one insists in viewing the Riemannian structure as  fundamental, then the physical situation described implies a point of non-differentiability. But with respect to the original smooth structure, everything is $C^\infty$ in the examples given.
\item[ii.] Two tetrads $e$ and $\tilde e$ related by a local Lorentz transformation define the same metric and are gauge-equivalent. Aren't then the field \eqref{uno}, and the Minkowski tetrad \eqref{mt}, which define the same metric, gauge-equivalent in this sense?  The answer is no, because a (proper orthochronus) Lorentz transformation does not change the direction of time.   Accordingly, the  quantity  detected by the fermion clock and not coded into the metric, is Lorentz-invariant. In fact, there are two signs which are not modified by a Lorentz transformation: the direction of time and the parity of space. These are modified, respectively, by Time reversal $T$ and Parity $P$.  Therefore there are \emph{two} signs that are invariant, corresponding to the four connected components of the full Lorentz group. These are captured for instance by the sign of the Lapse $n={\rm sign}(N)$ and the sign of the determinant of the tetrad, $s={\rm sign}(\det{e})=:n\,r$.  It is not difficult to see that fermion dynamics is affected by both. In this paper, for simplicity, we focus on the sign of the Lapse, and assume $r=+1$, mentioning the $r=-1$ only briefly. In this case, the sign $s$ of $\det{e}$ reflects the sign of the Lapse.
\end{itemize}

\section{Pockets}

\begin{figure}[h]
\begin{center}
\centerline{\includegraphics[scale=0.7]{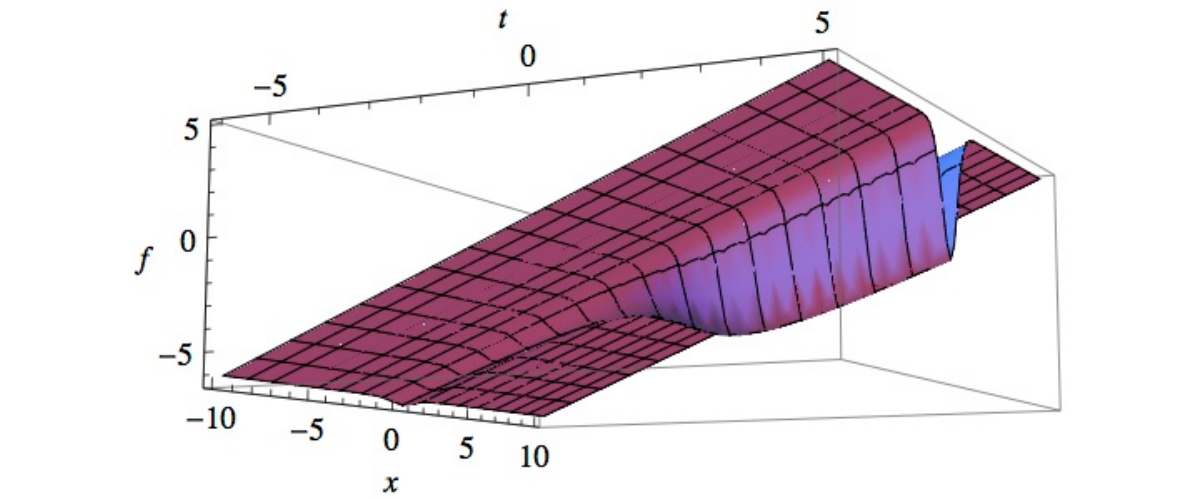}}
\caption{$f(x,t)$ (with $\tau\!=\!\sigma\!=\!\alpha\!=\!1)$. Its $t$-derivative is negative in a finite region.}
\label{graf}
\end{center}
\end{figure}

The negative-Lapse region of the gravitational field \eqref{uno} is extended all over spacetime.  We learn more about the physics of these regions by considering a region which is confined in space. As and example, consider the gravitational field 
\be
   e^0=d f\!(t,\vec x), \ \ \ \ \ \ \ \ e^i=dx^i  
   \label{ee}
\ee
where $f(t,\vec x)$ is a smooth function whose $t$ derivative is negative in some bounded region. An example is 
\be
   f(t,\vec x)=t-2\alpha \tau\  e^{-\frac{\vec x^2}{\sigma
   }}\left(\arctan({{t}/{\tau})}+{\pi}/{2}\right).
   \label{eee}
\ee
plotted in Fig.~\ref{graf}.  
\vspace{-2em}

\begin{figure}[h]
\begin{center}
\centerline{ \includegraphics[scale=0.4]{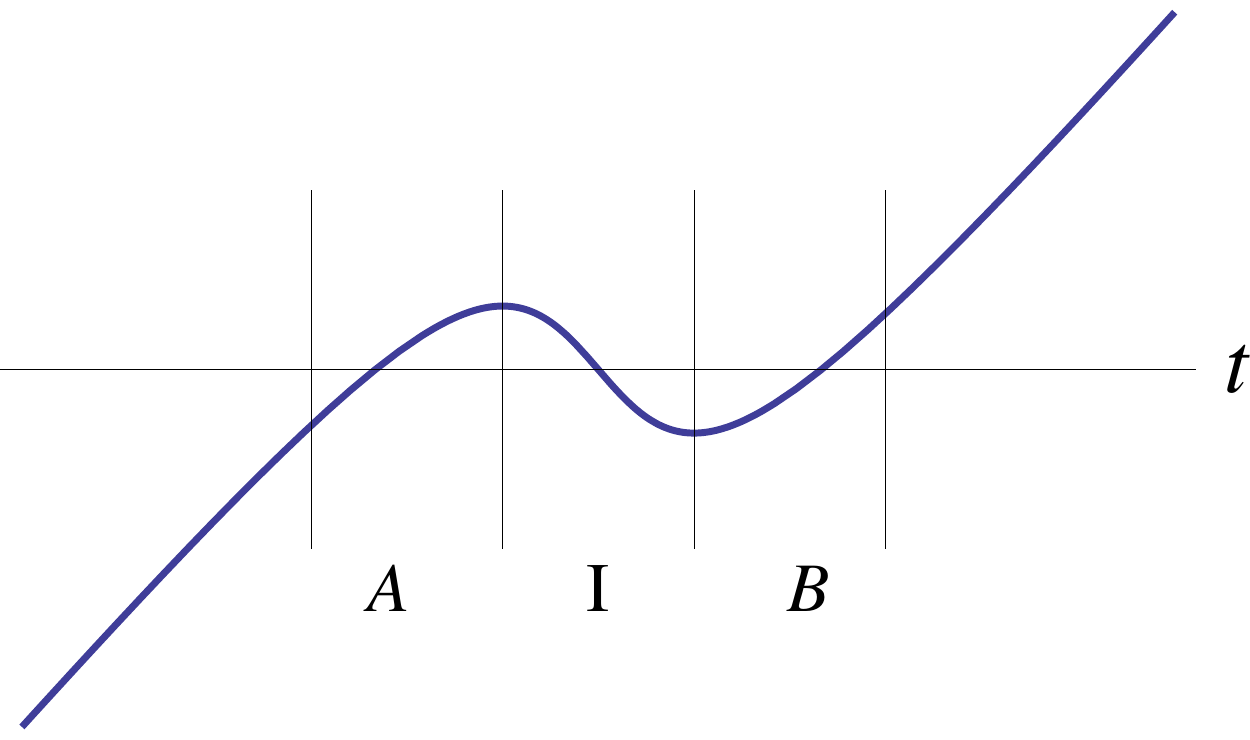}}
\caption{The three regions $A$, $I$ and $B$ where $f$ takes the same value ($\vec x=0$).}
\label{regioni}
\end{center}
\end{figure}

The field (\ref{ee}-\ref{eee})  reduces to (\ref{uno}-\ref{N}) for $\alpha\!=\!1$ and large $\sigma$. The determinant of the tetrad vanishes on the shell  
\be
   ({t}/{\tau})^2=2 \alpha \ e^{-\frac{\vec x^2}{\sigma}}-1. 
\ee 
and the Lapse is negative inside this shell.  Call ``region I"  the region inside the shell.  There are two other relevant regions: the two regions that surround the region I, characterized by the fact that the function $f$ takes the same values as in region I.  Call these region A and region B, respectively.  For instance, Fig.~\ref{regioni} illustrates the three regions for $\vec x=0$

\begin{figure}[h]
\begin{center}
\centerline{{ \includegraphics[scale=0.33]{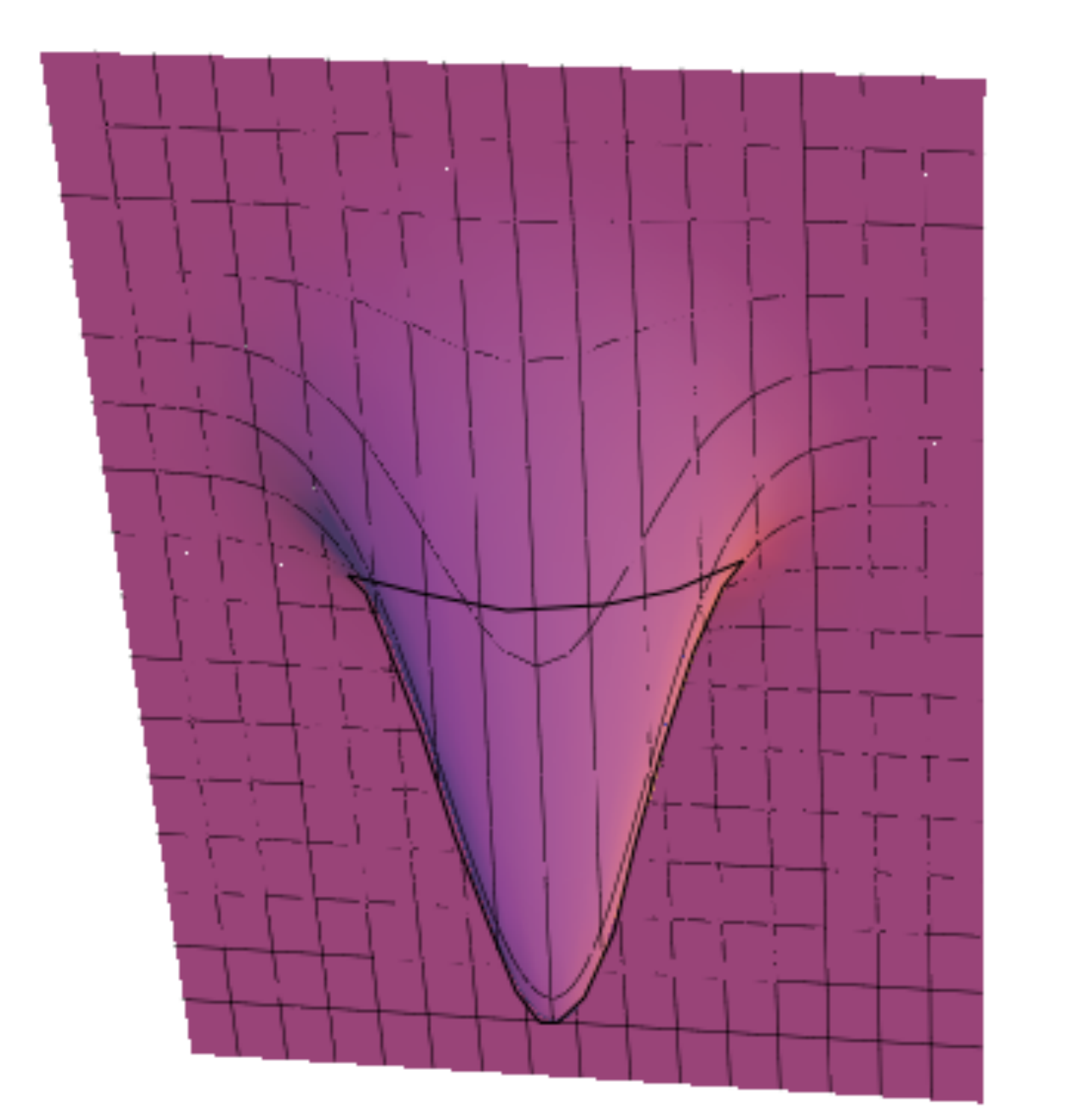}}{ \includegraphics[scale=0.27]{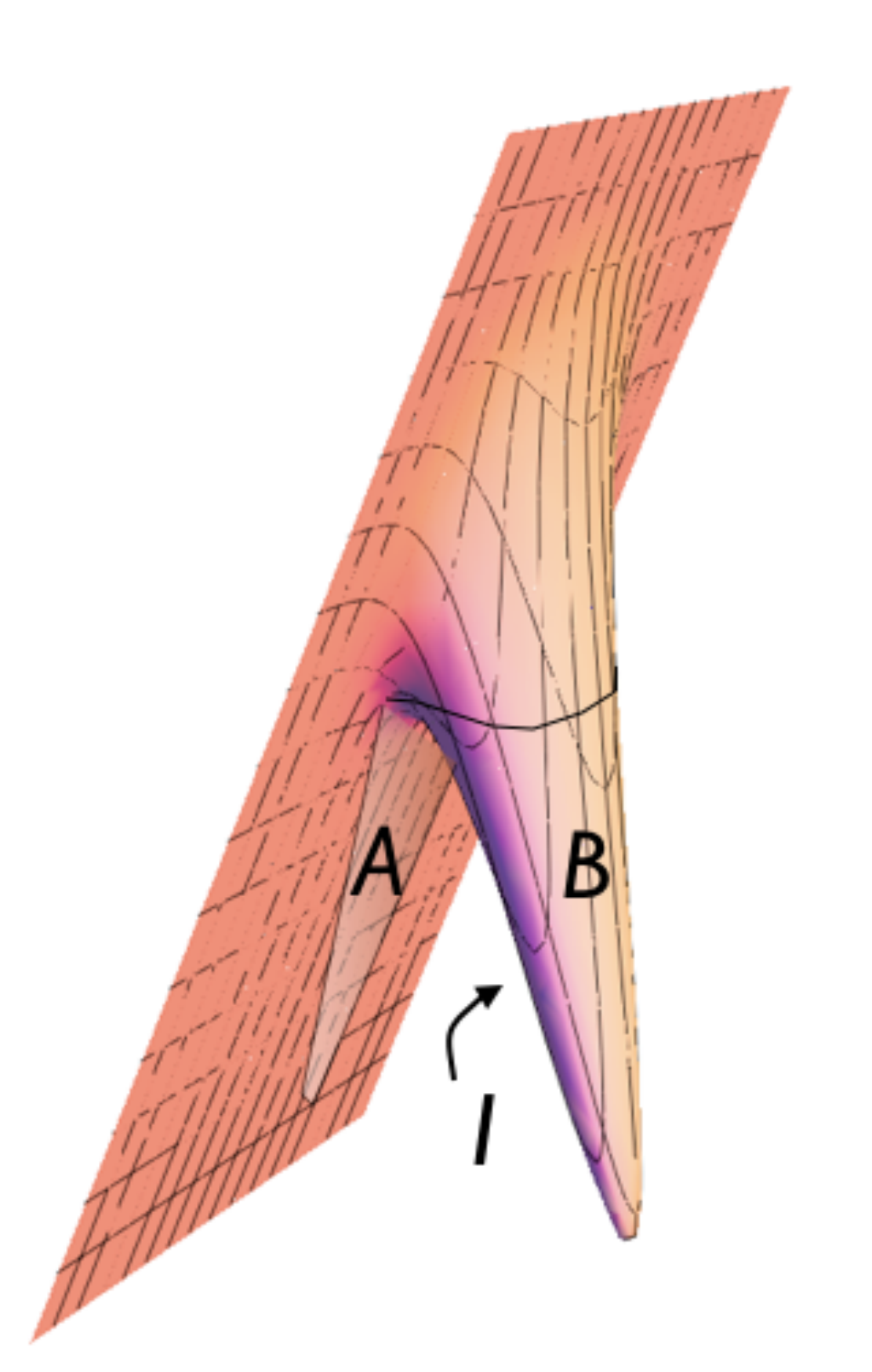}}}
\caption{A spacetime pocket (tetrad field \eqref{eee}), artificially immersed in an extra dimension for clarity, from two different perspectives. The Lapse is  always pointing upward in this representation.}
\label{tasca}
\end{center}
\end{figure}

The spin-connection of the field \eqref{ee} vanishes, therefore its spacetime has vanishing curvature: It follows that locally this spacetime is Minkowski space. But in this case spacetime is not globally Minkowski.  In fact, the map from this spacetime and Minkowski space in Minkowski coordinates $(T_M,\vec x)$ is given by 
\be
T_M = f(x,t)  \label{map}
\ee
But $f(x,t)$ is not injective. The three regions $A$, $B$ and $I$ are mapped onto the same region $R$ of Minkowski space by \eqref{map}.   The geometry defined by the gravitational field \eqref{ee} is that of a trouser's pocket: outside these three regions, spacetime is globally isomorphic to Minkowski space; each of the three regions is itself flat as well, but they are joined in such a way that $R$ is replaced by three regions with the same geometry as $R$: region $A$ is glued to the past, region $B$ is glued to the future, and region $I$ is glued to region $A$ and region $B$, as in Fig.~\ref{tasca}.
\vspace{-2em}

Now, the Dirac equation can be easily solved wit this gravitational field. A solution with four-momentum $p_I$ is
\be
   	\psi(\vec x,t)=e^{ip_I \int_Q^{(\vec x,t)}e^I} \ \psi_o
\ee
where the integral is a line integral from an arbitrary fixed spacetime point $Q$ to the point $(\vec x,t)$. The result does not depend on the path, because of the vanishing of $de^I$. For the field \eqref{ee}, this gives
\be
   	\psi(\vec x,t)=e^{ip_0f(\vec x, t)+ip_ix^i} \ \psi_o
\ee
We see immediately that the fermion dynamics outside the three regions is unaffected, since this is the standard solution in Minkowski coordinates: a solution on Minkowski space remains a solution on this field.  Futhermore, the solution inside the region $A$ is equal to the solution inside the region $B$ and is the time reversed of the solution inside the  negative-Lapse region $I$.  In other words, the fermion crosses the pocket as if only region $A$ existed. A pendulum at the origin, instead, would measure the time that takes into account the duration of all the three regions. 

Notice that causal behavior of the propagation is nontrivial: a fermion wave packet entering the region $A$ travels ahead in time in region $A$, then backward in time with respect to the sign of the Lapse in region $I$, and then again forward in time in region $B$ and since. This propagation is always consistent with the global causal structure defined by the metric.   

\begin{figure}[h]
\begin{center}
\centerline{{ \includegraphics[scale=0.25]{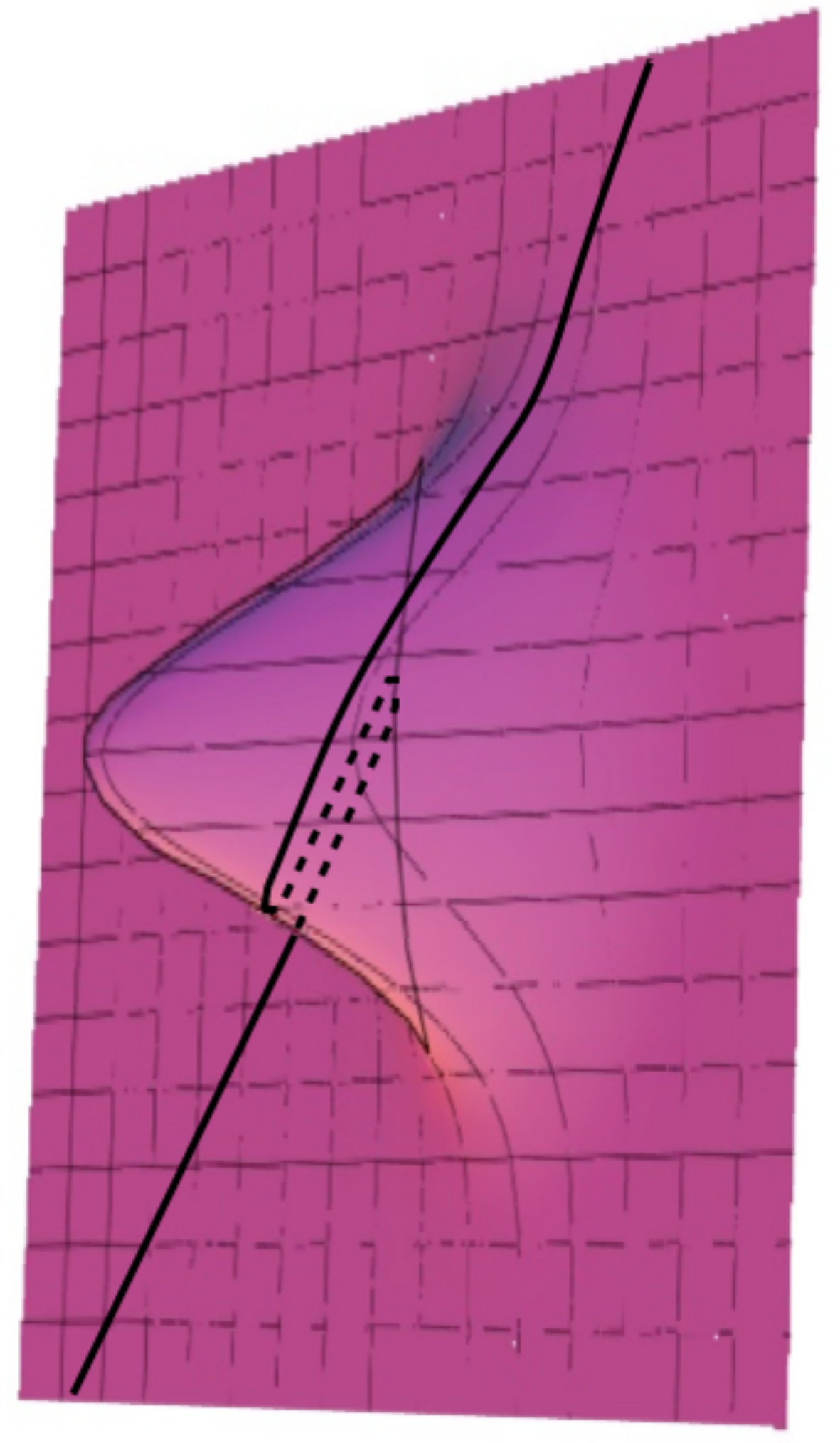}}}
\caption{An $r=-1$ pocket and the a-causal propagation of a wave packet.}
\label{tasca2}
\end{center}
\end{figure}

However, in the analogous situation where $r$ is negative the propagation of a wave packet violates the global causal structure defined by the metric. For instance, the field $e^0=dt, e^{1,2}=dx^{1,2}, e^3=df(z,t)$, where $f$ is non injective, determines an $r=-1$ pocket where a wave packet can follow the path represented in Fig.\ \ref{tasca2}. In region $I$ propagation is inconsistent with the causal structure defined by the metric. A fermion wave packet entering the region $A$ from the left of the picture would immediately give rise to two other fermion-wave-packets appearing inside the pocket: one traveling up the region $I$ and annihilating with the initial wave packet; the other traveling upward the region $B$ and emerging in the future. 

\vspace{-2em}

An objection to taking regions as the ones described as physically realistic comes from determinism. If the initial data are flat and the future is determined by the past, what could cause the development of a negative Lapse region? 

There are two possible answers. First, the flat case we have described is only a simple example. One might imagine that in a generic curved spacetime with matter the determinant of the metric can evolve towards zero, and the smooth continuation of the tetrad field evolution generate negative-$\det{e}$ regions.  More interestingly, pockets can appear in of-shell quantum fluctuations of the gravitational field, namely in the context of quantum gravity. In the next section, we turn to such a context.

In conclusion, if gravity is described by the tetrad, there might be spacetime regions where the determinant of $e$ is negative and a fermion clock runs backward with respect to a pendulum clock governed by the metric.  

Do such regions exist in nature, or should we assume that all gravitational fields in nature satisfy $N>0$ and $ \det{e}>0$?  Since the difference is measurable, the question is well posed.

\section{Quantum gravity and anti-spacetimes}

The question closing the previous Section becomes particularly significative in quantum gravity, when the dynamics is defined \`a la Feynman by summing over field configurations.  Should the sum be taken over all tetrad fields, or only over those with  $\det{e}>0$?  

If the gravitational action is taken to be the Einstein-Hilbert action $S_{EH}$ which depends on the metric, the issue is not much relevant, since tetrads with different signs of $\det{e}$ give the same action.  But the natural geometric action for the tetrad is not $S_{EH}$, but rather 
\be
S_{\text{tetrad}}=\frac{1}{4}\int {\rm tr} [e\wedge e\wedge F^*],
\ee
where $F$ is the curvature of the spin connection and the star indicates the Hodge dual in Minkowski space. This action  \emph{differs} from  $S_{EH}$ by a sign when $\det{e}<0$. This can be easily seen writing the two in tensor notation as functions of $e$:
\begin{eqnarray}
S_{EH}&=&\frac{1}{2}\int d^4x\;|\!\det e|\  R[e], \label{TB}\\
S_{\text{tetrad}}&=&\frac{1}{2}\int d^4x\; (\det e)\  R[e], \label{TA}
\end{eqnarray}
where $R$ is the Ricci scalar.  $S_{\text{tetrad}}$ defines a quantum theory where negative $\det e$ regions contribute to the amplitude with a negative sign. Loosely speaking, the metric is  the square of the tetrad, and a sign is lost in the metric formulation.  

The effect of this sign appears in covariant Loop Quantum Gravity (see \cite{Rovelli:2011eq} and references therein). The contribution of a single simplex to the dynamics is given in the semiclassical limit by a vertex amplitude satisfying \cite{Barrett:2009mw,Han:2011re} 
\be
     A \ \to \ e^{\frac{i}{\hbar} S}+e^{-\frac{i}{\hbar}S},
     \label{terms}
\ee
where $S$ is a discretization of $S_{EH}$ (with or without cosmological constant) associated to the simplex. The two terms in \eqref{terms} are determined by the two possible signs that $\det{e}$  can take on the simplex \cite{Barrett:2009mw,Han:2011re}.\footnote{Such terms play a role in quantum gravity amplitudes \cite{Bianchi:2010zs,Rovelli:2008dx} and may be sources of infrared divergences \cite{Aldo,Marios}. They are the sources of the characteristic ``spikes" of quantum Regge calculus.}    The same happens in the 3d Ponzano-Regge model \cite{Ponzano:1968uq,Barrett:2008wh}. 

If in nature it is always true that $\det{e}>0$, then the second terms in \eqref{terms} should perhaps be seen as spurious. Attempts to get rid of it have appeared in the literature \cite{Mikovic:2011bh,Neiman:2011gf,Engle:2012yg,Rovelli:2012yy}.  But if such regions can exist, then it is reasonable to assume that they affect the gravitational path integral, and the existence of the two terms in  \eqref{terms} simply reflects their existence. In other words, spacetime might fluctuate over negative $\det{e}$  regions.  The fact that these regions are physically distinguishable, as shown in this paper, renders this scenario more plausible. 

As noticed in  \cite{Rovelli:2012yy}, there is an analogy between negative $\det{e}$ regions and antiparticles. In both cases, the action contributes to the Feynman sum with a reversed sign. Both can be seen, in a sense, as configurations running backward in time.   For antiparticles, this is the intuition that grounds the St\"uckelberg-Feynman positron theory \cite{Stuckelberg,Feynman}, where antiparticles are interpreted as particles running backward in time.  The analogy is reinforced by the results of the previous sections, where we have shown that a fermion crossing a negative $\det{e}$ region is affected as if time was running backward therein. 

These observations justify the evocative name ``anti-spacetimes" suggested in \cite{Rovelli:2012yy} for such regions.  Feynman has given in \cite{FeynmanAntiP} a general physical argument indicating that antiparticles should exist as a general consequence of quantum theory and special relativity: positive-energy propagation necessarily spills outside the light cone; but spacelike propagation can be interpreted as backward in time in a different Lorentz frame.   Can an analogous argument be formulated in quantum gravity?

\section{Summary and open questions}

The existence of fermions indicates that the complete characterization of gravity requires a tetrad field $e_\mu^I(x)$. The natural geometric action for this field differs from the Einstein-Hilbert action by a sign factor, determined by the determinant  $\det{e}$. Regions of negative $\det{e}$, or ``anti-spacetimes", might  therefore contribute to the quantum gravity path integral.  This is what happens in the Ponzano-Regge model and in loop gravity, where the two signs of $\det{e}$ are at the source of the two terms in the asymptotic expansion \eqref{terms} of the vertex amplitude.  

Here we have shown that  if such  anti-spacetime regions exist, they are in principle detectable.  Over an ``anti-spacetime" region, the fermion's phase runs backward with respect to a standard clock. Tetrads corresponding to the same metric can be physically inequivalent. 

The are several possibilities:  (i) Anti-spacetimes do not exist, $\det{e}>0$ should constrain the gravity path integral, the difference between $S_{EH}$ and $S_{\text{tetrad}}$ is irrelevant.  (ii) Anti-spacetimes exist, but they contribute trivially to quantum gravity because the relevant action is $S_{EH}$. (iii)  The relevant action is $S_{\text{tetrad}}$, anti-spacetimes exist and contribute nontrivially to quantum gravity. We are inclined to think that the detectability of anti-spacetime, via fermion interference, enhances  the credibility of the last scenario.\footnote{But:  ``And though we should not have raised the argument unless we had thought it both important and likely to be answered ultimately in the affirmative, we still hold the opinion with some hesitation, with as much, indeed, as accompanies any conclusion we endeavor to draw respecting points in the very depths of science." \cite{faradayquote}}

Do anti-spacetimes exist in Nature?  Do they contribute to a Feynman formulation of quantum gravity in the same manner as antiparticles contribute to the total propagation of particles? These questions are well defined experimentally (``Are there regions where the phase of a fermion runs backward with respect to the Minkowski time?") and theoretically (``Which of the two actions \eqref{TA} and \eqref{TB} defines quantum gravity"?).

\end{document}